\documentclass[twocolumn]{jpsj3}

\usepackage[dvips]{color}
\usepackage{amsmath}

\title{Stability of FFLO states in optical lattices with bilayer structure}

\author{Yasuharu \textsc{Okawauchi} and \name{Akihisa \textsc{Koga}}
\thanks{E-mail address: koga@phys.titech.ac.jp}}

\inst{\address{Department of Physics, Tokyo Institute of Technology, 
Meguro, Tokyo 158-8551, Japan} }

\abst{ 
We investigate the stability of the superfluid state 
in a bilayer fermionic optical lattice system with a confining potential,
using the Bogoliubov de-Gennes equations.
It is clarified that in the imbalanced case, 
the introduction of the interlayer hopping 
stabilizes the radial Fulde-Ferrell-Larkin-Ovchinnikov (FFLO) state, 
while makes the angular FFLO state unstable.
We also discuss the system size dependence of the superfluid ground state.
It is clarified that in a certain ring region the A-FFLO state is indeed realized
in a large system.
}
\kword{FFLO state, Population imbalance, Bilayer optical lattice, 
Bogoliubov-de Gennes equation}

\begin{document}
\maketitle
\section{Introduction}
Recently ultracold atomic gases have attracted much interest
since the successful realization of Bose-Einstein condensation
(BEC) in a bosonic $^{87}$Rb system.\cite{Rb}
Among them, 
an ultracold gas system in a periodic potential, 
so-called, an optical lattice system,
\cite{BlochGreiner,Bloch,Jaksch,Morsch,Ketterle}
has been providing an ideal stage for experimental and
theoretical studies of fundamental problems in condensed
matter physics. 
Due to its high controllability in
interaction strength, particle number, and other parameters,
many remarkable phenomena have been observed such as
a phase transition between a Mott insulator and a superfluid
in bosonic systems,\cite{Greiner}
and a crossover between
the Bardeen-Cooper-Schrieffer (BCS) state and the BEC state
in fermionic systems.\cite{BCSBEC1,BCSBEC2,BCSBEC3}
In addition, the superfluid state in the spin-imbalanced fermionic systems
has been realized in a fermionic $^6$Li system,\cite{Imbalance1,Imbalance2}
which stimulates further theoretical investigations 
on the superfluid state and its related phenomena.

An interesting question for the fermionic optical lattice system
with imbalanced populations
is how the symmetry-breaking state is realized at low temperatures.
Various ground states have already been proposed to be more
stable than the polarized superfluid (PSF) state.\cite{DMFT1,DMFT2}
One of the probable candidates is 
the Fulde-Ferrell-Larkin-Ovchinnikov (FFLO) phase,
\cite{FF,LO} where Cooper pairs are formed with a finite total momentum. 
This phase has been observed in the high field region in $\rm CeCoIn_5$
\cite{CeCoIn1,CeCoIn2,CeCoIn3} and
has theoretically been discussed in the compound\cite{Adachi,Ikeda,YanaseCe,Miyake}
as well as cold atoms with imbalanced populations.
\cite{Mizushima0,Castorina0,Kinnunen0,DMRG,3DFFLO,Okumura,Liu}
In the two dimensional optical lattice with a confining potential,
it has been pointed out that two kinds of the FFLO states are realized such as
the radial-FFLO (R-FFLO)\cite{Mizushima,Iskin} and 
the angular-FFLO (A-FFLO) states.\cite{FFLO1,FFLO2}
In the former state, the superfluid order parameter changes its sign 
along the radial direction.
In the latter, 
the order parameter oscillates along the angular direction,
and the $C_{4v}$ symmetry as well as $U(1)$ symmetry are broken.
On the other hand, such FFLO states with a three-dimensional structure
have not been studied so well although the interlayer coupling should be
important for realistic optical lattice systems.\cite{2D_optical-lattice}

\begin{figure}[htb]
\begin{center}
\includegraphics[width=7.5cm]{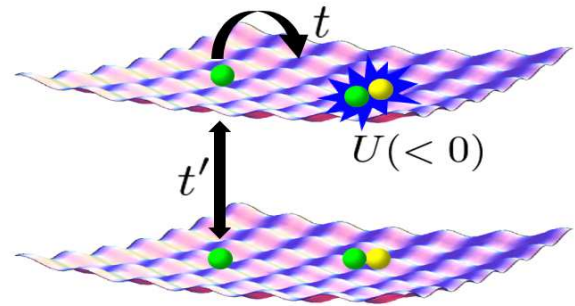}
\end{center}
\caption{(Color online) Optical lattice with a bilayer structure}
\label{fig:Bilayer optical lattice}
\end{figure}

To make this point clear, we systematically study 
the two-dimensional optical lattice system 
with a bilayer structure as a simple model,
which is schematically shown in Fig. \ref{fig:Bilayer optical lattice}.
We discuss how the interlayer hopping between two layers affects
ground state properties by means of
the Bogoliubov de-Gennes (BdG) equation.
We then clarify how the R-FFLO and A-FFLO states are realized 
in the optical lattice with a confining potential.
The system size dependence of the ground state is also discussed.

The paper is organized as follows. 
In \S\ref{sec:2}, we introduce the model Hamiltonian
for the two-component fermions in the optical lattice 
and briefly summarize our theoretical approach.
In \S\ref{sec:3}, we discuss the stability of 
the R- and A-FFLO states in the bilayer optical lattice system.
Scaling behavior in ground state properties is addressed in \S\ref{sec:4}.
A summary is given in the final section.

\section{Model and Method}\label{sec:2}
We study two-component ultracold fermions trapped in a two-dimensional 
bilayer optical lattice,
which should be described by the following Hubbard Hamiltonian,
\begin{equation}
\begin{split}
\hat{H}&=-t\sum_{\langle{ij}\rangle\sigma\alpha}\left(
\hat{c}^\dag_{i\alpha\sigma}\hat{c}_{j\alpha\sigma}+h.c.\right)\\
&-t'\sum_{i\sigma}\left(
\hat{c}^\dag_{i1\sigma}\hat{c}_{i2\sigma}+h.c.
\right)\\
&+\sum_{i\alpha\sigma}\left(V_i-\mu-h\sigma\right)\hat{n}_{i\alpha\sigma}
+U\sum_{i\alpha} \hat{n}_{i\alpha\uparrow}\hat{n}_{i\alpha\downarrow}
\label{eq.1}
\end{split}
\end{equation}
where $\langle{ij}\rangle$ indicates the nearest neighbors in each layer
with $L\times L$ sites.
$\hat{c}_{i\alpha\sigma}$ ($\hat{c}^\dag_{i\alpha\sigma}$) 
annihilates (creates) a fermion at the $i$th site of 
$\alpha(=1, 2)$th layer
with spin $\sigma$ (=$\uparrow$ , $\downarrow$), and
$\hat{n}_{i\alpha\sigma}={\hat c}_{i\alpha\sigma}^\dag {\hat c}_{i\alpha\sigma}$.
$t (t')$ is the intralayer (interlayer) hopping and 
$U(< 0)$ is the attractive interaction.
The total number of particles $N(=N_\uparrow + N_\downarrow)$ and 
the imbalanced population 
$P[=(N_{\uparrow}-N_{\downarrow})/(N_{\uparrow}+N_{\downarrow})]$,
where $N_\sigma=\sum_{i\alpha}\langle\hat{c}_{i\alpha\sigma}^\dag 
\hat{c}_{i\alpha\sigma}\rangle$,
are tuned by the chemical potential $\mu$ and the magnetic field $h$ 
although these quantities can be controlled directly in the experiments.
The confining potential is defined by 
$V_{i\alpha}=\beta (r_{i\alpha}/a)^2$, where 
$\beta (>0)$ is the curvature of the harmonic potential, $r_{i\alpha}$ 
is the distance measured from the center of the $\alpha$th layer, 
and $a$ is the lattice constant.

In the mean-field approach,
the interaction term should be given as
\begin{equation}
\begin{split}
U{\hat n}_{i\alpha\uparrow}{\hat n}_{i\alpha\downarrow}&\rightarrow 
U\sum_\sigma {\hat n}_{i\alpha\sigma}  n_{i\alpha\bar{\sigma}} 
-Un_{i\alpha\uparrow}n_{i\alpha\downarrow}\\
&+\Delta_{i\alpha}^*{\hat c}_{i\alpha\downarrow} {\hat c}_{i\alpha\uparrow} 
+\Delta_{i\alpha}{\hat c}_{i\alpha\uparrow}^\dag {\hat c}_{i\alpha\downarrow}^\dag 
-\frac{1}{U}|\Delta_{i\alpha}|^2,
\end{split}
\end{equation}
where $\Delta_{i\alpha}=
U\langle\hat{c}_{i\downarrow\alpha}\hat{c}_{i\uparrow\alpha}\rangle$ 
is the local superfluid order parameter, and 
$n_{i\sigma\alpha}=\langle\hat{c}^\dag_{i\sigma\alpha}
\hat{c}_{i\sigma\alpha}\rangle$ is the local particle density.
The BdG equations should be written 
in terms of the $4L^2\times 4L^2$ matrix, as
\begin{equation}
\sum_{j}\left(
\begin{array}{ccc}
{\bf H}^{(1)}_{ij}&{\bf T}_{ij}\\
{\bf T}_{ij}&{\bf H}^{(2)}_{ij}\\
\end{array}
\right)\left(\begin{array}{ccc}
{\bf {\phi}}^{(1)}_{qj}\\
{\bf {\phi}}^{(2)}_{qj}\\
\end{array}
\right)=
\epsilon_{q}
\left(
\begin{array}{ccc}
{\bf {\phi}}^{(1)}_{qi}\\
{\bf {\phi}}^{(2)}_{qi}\\
\end{array}
\right)
\label{eq.2}
\end{equation}
\begin{equation}
{\bf H}^{(\alpha)}_{ij}=
\left(\hspace{-3pt}
\begin{array}{ccc}
H^{\alpha}_{ij\uparrow}\hspace{-5pt}&\Delta_{ij}^{\ast}\\
{\Delta}_{ij}&-H^{\alpha}_{ij\downarrow}\\
\end{array}
\hspace{-6pt}\right)
\label{eq.3}
\end{equation}
\begin{equation}
{\bf T}_{ij}=
\left(
\begin{array}{ccc}
K_{ij}&0\\
0&-K_{ij}\\
\end{array}
\right)
\label{eq.4}
\end{equation}
where $H^{\alpha}_{ij\sigma}=-t\delta_{\langle ij\rangle}+
(-\mu-h\sigma+Un_{i\alpha\bar{\sigma}})\delta_{ij}$,
$K_{ij}=-t'\delta_{ij}$, and $\Delta_{ij}={\Delta}_{i}\delta_{ij}$,
where $\delta_{ij}$ is the Kronecker delta. 
The eigenfunctions $\phi^{(\alpha)}_{qi}={}^t(u^{(\alpha)}_{qi\uparrow},
v^{(\alpha)}_{qi\downarrow})$ indicate 
the Bogoliubov quasiparticle amplitudes at the $i$th site on the $\alpha$th layer.
The mean fields are then determined by the self-consistent equations as
\begin{eqnarray}
\Delta_{i\alpha}&=&
U\sum_{q}u^{(\alpha)}_{qi}v^{(\alpha)}_{qi}f(\epsilon_{q}), \\
n_{i\alpha\uparrow}&=&\sum_{q}\left|u^{(\alpha)}_{qi}
\right|^{2}f(\epsilon_{q}), \\
n_{i\alpha\downarrow}&=&\sum_{q}\left|v^{(\alpha)}_{qi}
\right|^{2}f(-\epsilon_{q})
\end{eqnarray}
where $f(\epsilon)$ = $[\exp(\epsilon/T) +1]^{-1}$ 
is the Fermi distribution function and $T$ is the temperature. 

In the iterative method, one sometimes reaches 
the metastable state with the higher energy than the ground state.
Therefore, it is important to choose an appropriate initial state 
in this treatment.
In this paper, to discuss the stability of the FFLO states, 
we use the symmetry breaking states with $n$-fold symmetry in space
and solve the BdG equations iteratively.
Then we determine the ground state by comparing the free energies 
of their converged solutions.

For convenience, 
we define the characteristic length of the potential 
as $d=\sqrt{t/\beta}a$ and the effective particle density as
$\tilde{\rho}=Na^2/\pi d^2$.
We set $t$ as a unit of energy and fix the interaction strength and the temperature 
as $U/t=-4$ and $T/t=0.001$, for simplicity.
The temperature is low enough, which enables us to discuss ground state properties 
in the bilayer optical lattice system.
The obtained ground state always has a mirror symmetry
in the case $t'/t\neq 0$.
Therefore, in this paper,
we will show the results for one of the layers
to discuss the stability of the superfluid state.

\section{Stability of the superfluid state}\label{sec:3}
In the section, we focus on the optical lattice system 
($L=30$) 
with the confining potential 
$\beta=0.04$ $(d=5a)$ to discuss how 
the introduction of the interlayer hopping affects the stability 
of the BCS, R-FFLO and A-FFLO states.
Before starting discussions, 
we would like to mention the characteristic profiles for these states.
When the magnetic field is small enough,
the BCS state is realized, where the finite 
pair potential appears without the magnetization.
In the R-FFLO or A-FFLO states, 
the magnetization is finite, and
the sign changes appear in the pair potential,
depending on the direction 
of its oscillation.
Namely, the sign change in the radial direction appears in
the former state. 
On the other hand, it appears in a certain ring region in the latter state.
Note that the BCS and R-FFLO states have the same spatial symmetry,
in contrast to the A-FFLO state.
Therefore, the BCS state is adiabatically connected to the R-FFLO state 
through the crossover.

\begin{figure}[htb]
  \begin{center}
  \includegraphics[width=8cm]{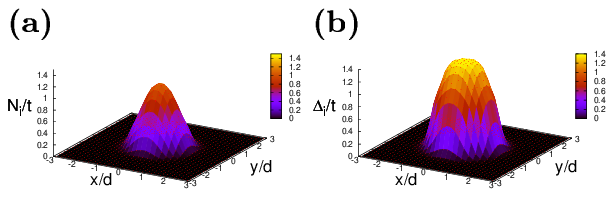}
 \end{center}
  \caption{Profiles of the particle density $N_i$ and 
  the order parameter $\Delta_{i}$ in the dilute system 
with $\tilde{\rho} \simeq 0.89$, $t'/t=0.0$ and $P = 0$.}
  \label{fig:den1}
\end{figure}

\begin{figure}[htb]
\begin{center}
\includegraphics[width=6.5cm]{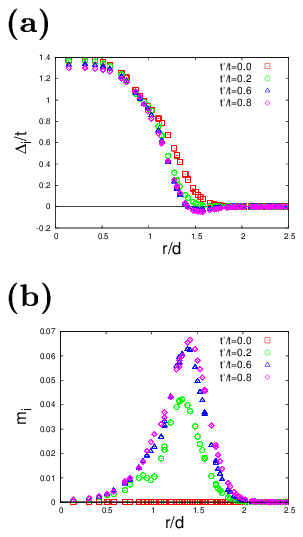}
\end{center}
\caption{(Color online) Profile of the order parameter $\Delta_i$ (a) and 
the magnetization $m_i$ (b) as a function 
of $r/d$ in the dilute system.}
\label{fig:0.6}
\end{figure}

We first study ground state properties of the dilute system with 
$\tilde{\rho} \simeq 0.89$ $(N\simeq 70)$.
In the case with $t'/t=h/t=0$, the particles are smoothly distributed 
and the order parameter appears around the center of the system.
The local particle density $N_i=\sum_\sigma n_{i\alpha\sigma}$ 
and local order parameter $\Delta_i$ for each layer 
are shown in Fig. \ref{fig:den1}. 
When the magnetic field is applied,
three kinds of ground states appear.
In the case $h<h' (\sim 0.7t)$, the BCS state is realized.
In the intermediate region $h' < h < h_c(\sim 1.4 t)$,
the R-FFLO state is realized, where
the magnetization appears in the system and 
the order parameter changes its sign along the radial direction.
Beyond the critical field $h_c$, 
the pair potential is no longer finite, and the ground state is paramagnetic.
Here we focus on the system with $h/t=0.6$ to discuss the effect of 
the interlayer hopping.
The cross-sections of the order parameter and the magnetization,
which is defined by $m_i=n_{i\alpha\uparrow}-n_{i\alpha\downarrow}$,
are shown in Fig. \ref{fig:0.6}.
When the interlayer hopping $t'$ is small, 
the system belongs to the BCS state,
where the order parameter appears around the center of the potential
($r/d < 1.7$).
The introduction of the interlayer hopping reduces the pair potential,
typically around $r/d \sim 1.5$. 
At last, the sign change appears in the superfluid order parameter
although it may not be visible in the case $t'/t=0.2$ 
in Fig.~\ref{fig:0.6}~(a).
In addition, the magnetization is induced in the vicinity of the sign change point,
as shown in Fig.~\ref{fig:0.6}~(b).
This suggests the realization of the R-FFLO state.
Further increase in the interlayer hopping
yields a clear sign change in the order parameter and
increases the magnetization monotonically.
Therefore, we can say that the interlayer hopping stabilizes the R-FFLO state.

\begin{figure}[htb]
\begin{center}
\includegraphics[width=7cm]{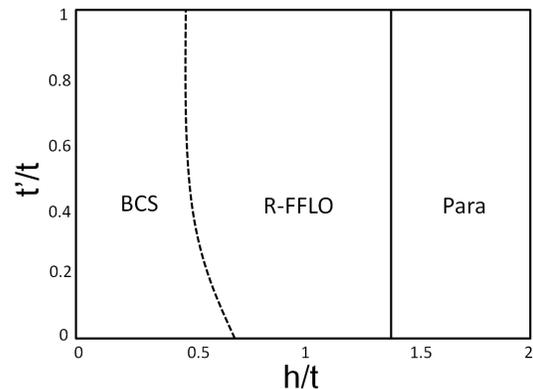}
\end{center}
\caption{Phase diagram of the dilute system with $\tilde{\rho}\simeq 0.89$}.
\label{fig:phase_dilute}
\end{figure}

By performing similar calculations, 
we obtain the phase diagram of the dilute system,
as shown in Fig. \ref{fig:phase_dilute}.
The crossover between the BCS state and the R-FFLO state
is roughly estimated by the appearance of the magnetization,
which is shown as a dashed line.
The interlayer hopping $t'$ little affects the phase boundary between 
the R-FFLO and normal states, in contrast to the crossover boundary.

\begin{figure}[htb]
  \begin{center}
  \includegraphics[width=8cm]{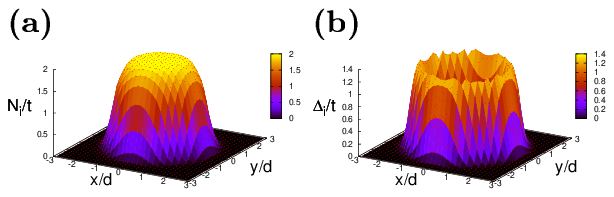}
 \end{center}
  \caption{Profiles of the particle density $N_i$ and 
  the order parameter $\Delta_{i}$ in the dense system 
with $\tilde{\rho} \simeq 6.4$, $t'/t=0.0$ and $P = 0$.}
  \label{fig:den2}
\end{figure}

\begin{figure}[htb]
  \begin{center}
  \includegraphics[width=8cm]{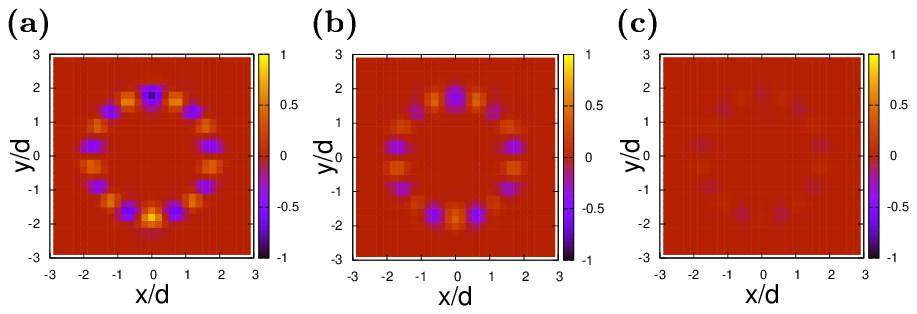}
  \end{center}
  \caption{(Color online) 
  Order parameter $\Delta_{i}$ in the dense system 
when $t'/t=0.0, 0.6$ and $1.0$ (from the left to the right).}
\label{fig:hop}
\end{figure}

Next, we study ground state properties of the dense system with 
$\tilde{\rho} \simeq 6.4$ ($N\simeq 500$).
In the case, a doubly occupied (Fock) state is realized 
around the center of the system 
due to the attractive interaction and the trap potential,
as shown in Fig. \ref{fig:den2} (a).
Therefore, when $h/t=0.0$, the superfluid state is realized away from the center
 and the doughnutlike structure appears in the pair potential,
 as shown in Fig. \ref{fig:den2} (b). 
When the spin imbalanced population is introduced, 
the R-FFLO and A-FFLO states are realized 
in the cases $(h'< h <h_{c1})$ and $(h_{c1}< h <h_{c2})$, 
where $h'=0.6t$, $h_{c1}=1.1t$ and $h_{c2}=1.4t$.
We here focus on the latter case with $h/t=1.3$,
where oscillation behavior with nine peaks in the order parameter appear 
in a certain ring region, 
as shown in Fig. \ref{fig:hop} (a). 
The introduction of the interlayer hopping $t'$ monotonically decreases
the magnitude of the order parameters, as shown in Fig. \ref{fig:hop}.
Finally it vanishes and the phase transition occurs to the paramagnetic state.
This instability should be explained by the following.
In the A-FFLO state,
the oscillation of the order parameter appears in a certain ring region,
which implies that 
one dimensional structure plays an important role to be stabilized.
In this point of view, 
the interlayer coupling can be regarded as the introduction of the 
two-dimensional structure.
Therefore, the A-FFLO state becomes unstable 
against the interlayer coupling.

\begin{figure}[htb]
\begin{center}
\includegraphics[width=8cm]{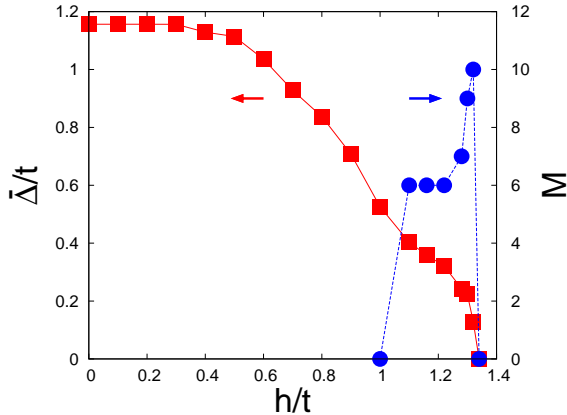}
\end{center}
\caption{(Color online)
Squares and circles represent the average and
the number of peaks in the angular direction of the order parameter 
in the dense system with $t'/t=0.6$.}
\label{fig:t=0.6_AFFLO}
\end{figure}

To discuss how the magnetic field affects the nature of the superfluid state,
we also show the number of peaks in the angular direction $M$ and 
the average of the order parameter,
which is defined by $\bar{\Delta}=\sum_i|\Delta_i|/N_\downarrow$,
in Fig. \ref{fig:t=0.6_AFFLO}.
When $h/t=0.0$, the BCS state is realized with $\bar{\Delta}/t\sim 1.2$ and $M=0$.
Applying the magnetic field, $\bar{\Delta}$ is little changed 
up to $h=h'(\sim 0.3t)$
and begins to decrease beyond it.
This means that the crossover occurs to the R-FFLO state at $h=h'$,
where the sign change in the order parameter reduces $\bar{\Delta}$. 
Further increase in the magnetic field monotonically decreases the quantity.
When $h=h_{c1}$, oscillation behavior suddenly appears in the angular direction of 
the order parameter $(M=6)$,
and the phase transition occurs to the A-FFLO state.
The R-FFLO and A-FFLO states have different spatial structures
in the magnetization and order parameter. For example, 
the sign change in the order parameter appears in the radial and angular direction,
which means that these states
are not adiabatically connected to each other around $h=h_{c1}$.
This is contrast to the results for one-dimensional system, where
the second-order phase transition occurs 
between the BCS and FFLO states.\cite{Machida0}
When the system approaches the phase boundary ($h_{c2}\sim 1.4t$), 
$\bar{\Delta}$ is rapidly decreased and $M$ is increased. 
This may originate from the fact that
the magnetization can be induced around the regions with $\Delta_{i\alpha}/t\sim 0$.
Therefore, when the large magnetic field is applied, 
many peaks appear in the profile of the order parameters.
This tendency is essentially the same as 
the single layer optical lattice.\cite{FFLO1}
Finally, the order parameter vanishes at $h=h_{c2}$, where
the phase transition occurs to the normal state.

\begin{figure}[htb]
\begin{center}
\includegraphics[width=7cm]{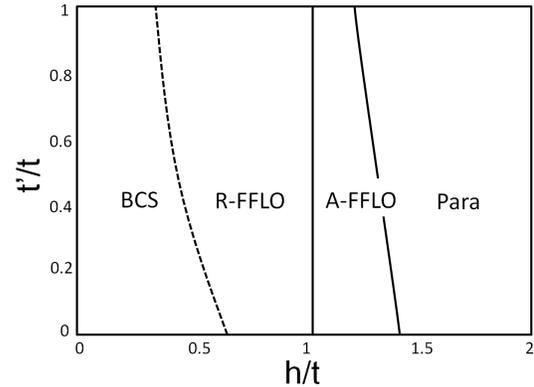}
\end{center}
\caption{Phase diagram for the dense system with 
$\tilde{\rho} \simeq 6.4$}.
\label{fig:t-h_N=500}
\end{figure}

By performing the similar calculations, we obtain the phase diagram
shown in Fig. \ref{fig:t-h_N=500}.
It is found that the interlayer hopping stabilizes the R-FFLO state,
which is the same as the dilute case.
On the other hand, the A-FFLO state becomes unstable
against the interlayer hopping.

We have treated the bilayer system to discuss how the interlayer hopping affects
the stability of the FFLO states.
Although the treated model is simple to discuss ground state properties of
the layered optical lattice system, we believe that
the obtained results capture the essence of the interlayer hopping. Namely,
the A-FFLO state with interesting spatial properties survives 
in the system with a small interlayer hopping.
In the following section, we discuss how the FFLO states are realized 
in the large system.

\section{Size dependence of FFLO states}\label{sec:4}
In the section, we study the stability of the FFLO states in the larger system.
It is naively expected that low energy properties in the system are scaled by 
the characteristic length of the harmonic potential.
By contrast, the characteristic length of the FFLO state 
should depend on the spin imbalance in the system, which 
is not directly related to the length $d$,
as discussed in the previous section.
Therefore, it is necessary to
clarify how interesting low energy properties are 
changed by the system size.

\begin{figure}[htb]
  \begin{center}
  \includegraphics[width=8cm]{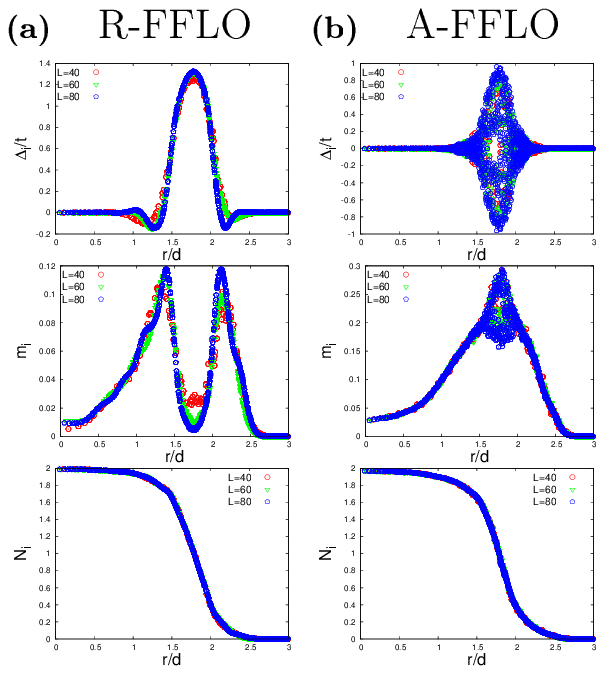}
 \end{center}
  \caption{(Color online) 
Profiles of the order parameter $\Delta_i$, magnetization $m_i$, 
and particle density $N_i$
for the R-FFLO (a) and A-FFLO (b) states.}
  \label{fig:FFLO}
\end{figure}

\begin{figure}[htb]
  \begin{center}
  \includegraphics[width=8cm]{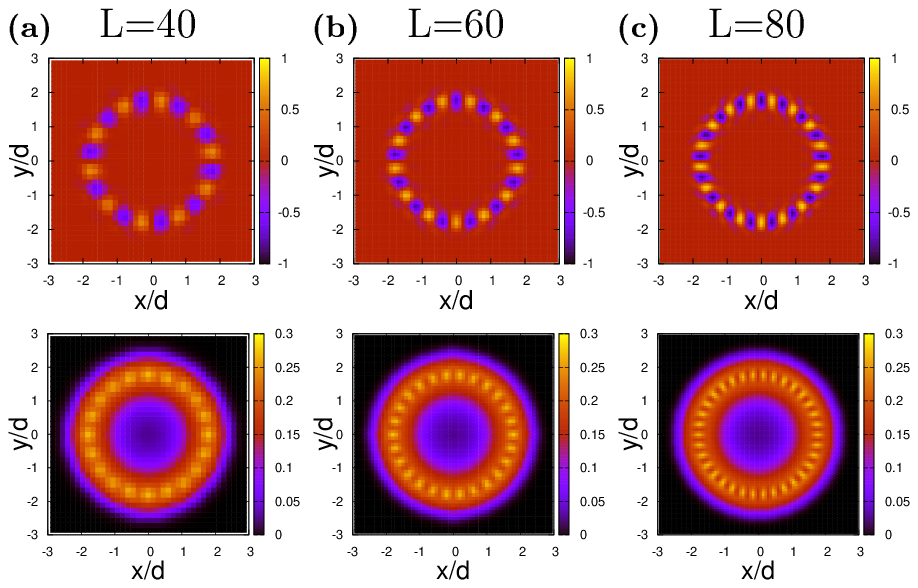}
 \end{center}
  \caption{Order parameter (upper panel) and magnetization (lower panel) 
  in the system with $t'/t=0.6$, $\tilde{\rho}\simeq 6.4$, $P\simeq0.15$, 
  $L=40, 60$, and $80$ (from the left to the right).}
  \label{fig:Density-FFLO}
\end{figure}

\begin{figure}[htb]
\begin{center}
\includegraphics[width=8cm]{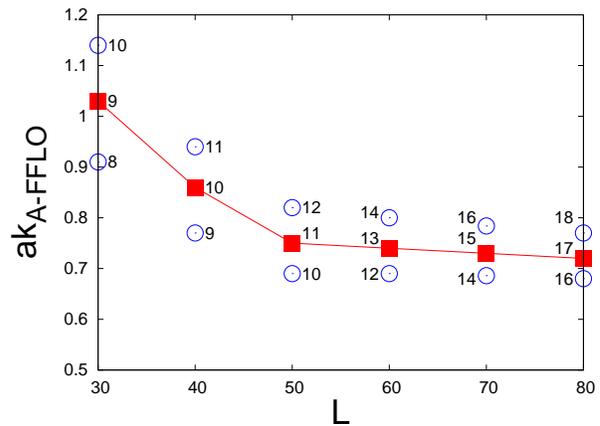}
\end{center}
\caption{
Squares (circles) represent the wave number characteristic of the A-FFLO state 
for the ground (meta-stable) state.
The number indicates peaks in the pair potential.
}
\label{fig:L-k}
\end{figure}

To discuss scaling behavior in the system carefully,
we fix the confining potential 
at four edges of the system as $V_{i\alpha}=18t$,
and keep the effective particle density and
the spin imbalance as $\tilde{\rho}=6.4$ and $P=0.15$.
The cross-sections of the particle density and the order parameter 
are obtained, as shown in Fig. \ref{fig:FFLO}.
In the BCS case (not shown), 
the profiles are well scaled,
and thereby the state is stable in the large system.
We find in Fig. \ref{fig:FFLO} (a) that in the R-FFLO state, 
a main peak structure in the order parameter is scaled.
However, a size dependence appears around the regions 
with a sign change ($r/d\sim 1.3$ and 2.2).
Since the dip structures tend to shrink on the increase of the system size,
it may be difficult to observe them in a large system,
as a signature for the realization of the R-FFLO state.
As for the A-FFLO state, 
disorder behavior in the profiles of the order parameter and magnetization 
appears in the region $(1.5<r/d<2.0)$, which is reflected by the oscillation 
of these quantities in the angular direction, as shown in Fig. \ref{fig:FFLO} (b).
Nevertheless, their envelopes are well scaled, where
the magnitude has a maximum at $r_0/d=1.75$.
The spatial dependence of the order parameter and magnetization is also shown 
in Fig. \ref{fig:Density-FFLO}.
It is clearly found that each peak in 
the magnetization $m_{i}$ is located at the sign change point
along a certain ring in the order parameter $\Delta_{i}$. 
These are consistent with the fact 
that the A-FFLO state is realized in the ring region.
Increasing the system size, we find that the number of peaks 
in a certain ring region is increased.
This means that the peak structure in the angular direction is not scaled.
To clarify the size dependence of the oscillations, 
we also calculate the wave number defined by 
$k_{\rm{A-FFLO}}=M/r_0$,
where $M$ is the number of the peaks in the order parameter.
Fig. \ref{fig:L-k} shows that it approaches  
a certain value $k_{\rm{A-FFLO}}\sim 0.7/a$ when the system size is increased.
These imply that there indeed exists a region with the A-FFLO state,
and it is connected to the FFLO state expected in the one-dimensional Fermi gases.
The characteristic wave number does not depend on the system size, 
but on the imbalanced populations.
Therefore, we can say that 
the A-FFLO state should be realized in the thermodynamic limit 
$(L\rightarrow\infty)$ if the interlayer hopping is small enough.

Before closing this section, 
we wish to comment on the possibility of the supersolid state.
In the attractive Hubbard model on the bipartite lattice 
in two or higher dimensions,
it is known that 
the density wave ground state and the superfluid state are degenerate
at half filling.\cite{Shiba,Scalettar,Freericks}
Therefore, in a certain narrow region with $\langle n_i \rangle \sim 1$,
the supersolid state, where both the density wave state 
and superfluid state coexist, 
may be realized.\cite{Supersolid}
It is necessary to carefully deal with particle correlations
beyond the BdG mean-field approach, which is now under considerations.

\section{Summary}
We have studied the stability of the superfluid state 
in a bilayer fermionic optical lattice system with imbalanced populations
by means of the BdG equations.
It has been clarified that the introduction of the hopping between two layers 
stabilizes the radial FFLO state, while makes the angular FFLO state unstable.
We have also discussed scaling behavior in the superfluid state.
It has been found that the wave number characteristic of the A-FFLO state 
approaches a certain value when the system size is increased.
This implies that the A-FFLO state should be realized 
in the thermodynamic limit.

\section*{Acknowledgment}
The authors thank A. Rosch, M. Tezuka, S. Tsuchiya, and Y. Yanase, 
for valuable discussions. 
This work was partly supported by the Grant-in-Aid for Scientific Research 
20740194 (A.K.) and 
the Global COE Program ``Nanoscience and Quantum Physics" from 
the Ministry of Education, Culture, Sports, Science and Technology (MEXT) 
of Japan.

\end{document}